\documentclass[sn-nature]{sn-jnl}

\usepackage{graphicx}%
\usepackage{multirow}%
\usepackage{amsmath,amssymb,amsfonts}%
\usepackage{amsthm}%
\usepackage{mathrsfs}%
\usepackage[title]{appendix}%
\usepackage{xcolor}%
\usepackage{textcomp}%
\usepackage{manyfoot}%
\usepackage{booktabs}%
\usepackage{algorithm}%
\usepackage{algorithmicx}%
\usepackage{algpseudocode}%
\usepackage{listings}%
\usepackage{soul}

\theoremstyle{thmstyleone}%
\theoremstyle{thmstyletwo}%

\theoremstyle{thmstylethree}%

\begin{document}

\title[Article Title]{Perspectives on the Physics {\color{black}of Late-Type Stars} from {\color{black}Beyond} Low Earth Orbit, {\color{black}the Moon and Mars}} 


\author*[1,2]{\fnm{Savita} \sur{Mathur}}\email{smathur@iac.es}

\author[3]{\fnm{\^Angela R. G.} \sur{Santos}}\email{Angela.Santos@astro.up.pt}
\equalcont{These authors contributed equally to this work.}


\affil*[1]{\orgname{Instituto de Astrof\'\i sica de Canarias (IAC)}, \orgaddress{\postcode{E-38205}, \city{La Laguna}, \state{Tenerife}, \country{Spain}}}

\affil[2]{\orgdiv{Departamento de Astrof\'isica}, \orgname{Universidad de La Laguna (ULL)}, \orgaddress{\postcode{E-38206}, \city{La Laguna},  \state{Tenerife}, \country{Spain}}}

\affil[3]{\orgdiv{Instituto de Astrof\'isica e Ci\^encias do Espa\c{c}o, Universidade do Porto}, \orgname{Universidade do Porto, CAUP}, \orgaddress{\street{Rua das Estrelas},  \postcode{PT4150-762}, \city{Porto}, \country{Portugal}}}


\abstract{With the new discoveries enabled thanks to the recent space missions, stellar physics is going through a revolution. However, these discoveries opened the door to many new questions that require more observations. The European Space Agency’s Human and Robotic Exploration programme provides an excellent opportunity to push forward the limits of our knowledge and better understand stellar structure and dynamics evolution. Long-term observations, Ultra-Violet observations, and a stellar imager are a few highlights of proposed missions {\color{black} for late-type stars} that will enhance the already planned space missions.}




\maketitle

\section{Introduction}\label{sec1}

Recently stellar physics has gone through a revolution. Interestingly, a large number of exciting discoveries have been made by missions whose main goal was not to study stars themselves, but rather to detect exoplanets around them. In particular,  a lot of progress has been reached thanks to the high-quality data collected by space telescopes accurately measuring stellar brightness to detect planetary transits in front of the host star, such as ESA {\color{black}(European Space Agency)} CoRoT  (Convection, Rotation, and Transits) \citep[][]{2006ESASP.624E..34B}, the NASA (National Aeronautics and Space Administration) Kepler/K2 mission \citep{2010Sci...327..977B,2014PASP..126..398H}, and now the NASA TESS mission (Transiting Exoplanets Survey Satellite) \citep{2015JATIS...1a4003R}. For example, they provided new insights into the nature of stellar magnetic activity revealing the connection between stellar photometric variability, rotational period and temperature \citep[see, e.g.][]{2014ApJS..211...24M,2019A&A...621A..21R,2019ApJS..244...21S,2021ApJS..255...17S}, led to the discoveries of superflares in solar-like stars \citep{Maehara2012, Okamoto2021} and even allowed probing deep layers of stars thanks to asteroseismology  \citep{2019LRSP...16....4G,2021RvMP...93a5001A,2022ARA&A..60...31K}. They also {\color{black} opened} a debate on how stars spin down as they age and on how to connect stellar ages with rotation and magnetic activity \citep{2016Natur.529..181V, 2016ApJ...823...16B, Curtis2019, 2023A&A...672A.159G,2023ApJ...952..131M}. Studies of stellar magnetic activity go in hand with solar physics since they allow {\color{black} us to put} the Sun in the context of other stars \citep{2020Sci...368..518R, 2023A&A...672A..56S}, {\color{black} and to} understand its magnetic past and future \citep{2018A&A...619A..73L}, and the resulting environment in the solar system. Indeed, the development of life and habitability can be affected by the magnetism of the host stars, which can contribute to the loss of their atmospheres and change the architecture of the systems \citep[e.g.][]{2017ARA&A..55..433K,2019A&A...630A..52R,Elsaesser2023}. In addition, magnetic activity and rotation can hamper planet detection \citep[e.g.][]{2013A&A...556A..19O}.




Another highlight concerns evolved solar-like stars (subgiants and red giants), where thanks to the study of mixed modes, that behave as acoustic modes in stellar envelopes and as gravity modes in the radiative interiors, we can probe the core of those stars \citep[e.g.][]{2011Sci...332..205B}. By studying the pattern of these mixed modes in red giants, we are now able to distinguish between a red giant burning H in a shell and a red giant burning He in the core \citep[e.g.][]{2011Natur.471..608B}. We can also retrieve information on the stellar rotation in the core of red giants and subgiants \citep[e.g.][]{2012Natur.481...55B,2014A&A...564A..27D,2018A&A...616A..24G}. While these are extremely important progresses that were made in the last decade, they open even more questions about the evolution of those stars, in particular on the transport of angular momentum. The current models do not reproduce the observed rates in the cores of subgiants and red giants, highlighting the need to improve the theoretical models though some progress is being made \citep[e.g.][]{2013A&A...555A..54C, 2022A&A...664L..16E}. 


Stars are the building blocks of the universe. Their precise understanding and characterization is key for the study of the evolution of planetary systems and of our Galaxy (also known as galacto-archeology). Asteroseismology being able to provide stellar fundamental parameters with a high precision is invaluable for these other fields. But to reach such precision on mass, radius, and age, we need to be able to detect individual modes, which requires observations as long as possible with as few gaps as possible in the data.  

{\color{black}F}uture missions led by ESA, PLATO (PLAnetary Transits and Oscillations of stars; scheduled for the end of 2026) \citep{2014ExA....38..249R},  and NASA, Roman Space Telescope \citep{2023ApJS..269....5W}, will provide data crucial to advance the field in many of these aspects{\color{black}. Nevertheless,} the ESA Human and Robotic Exploration (HRE) would be a great opportunity to go beyond with EUV (extreme ultraviolet) or {\color{black}develop} stellar imager instruments for {\color{black} Beyond Low-Earth Orbit (BLEO), the Moon and/or Mars} experiments that are not part of any future selected mission so far.

Below, we will first go through the current key knowledge gaps in stellar physics, then discuss the priority for future space programs, and finish with a summary.

\section{Key knowledge gaps}

Previous space missions such as CoRoT, {\it Kepler}/K2, and now TESS showed and are still showing the power of asteroseismic analyses. While providing new insights into stellar evolution, internal structure, and dynamics of stars from the main sequence {\color{black} (MS)} to the red giants {\color{black} (RG)}, these observations opened new questions. We focus here on key knowledge gaps in {\color{black}late-type} stars.

\subsection{Core-envelope coupling in solar-like stars}\label{sec:rotgap}

One of the unexpected findings prompted by \textit{Kepler} was the bimodality of the rotation-period distribution for stars in the unsaturated regime \citep{2013MNRAS.432.1203M,2014ApJS..211...24M}. This bimodality results in a gap at intermediate rotation and is now known to be related to stellar evolution. Using clusters, for which ages can be accurately determined, the rotation evolution of K stars was found to stall between ages of 0.6 and 1 Gyr \citep{Curtis2019}. While the rotation sequence of young clusters is monotonic with stellar mass, this behaviour is not found for old clusters: their rotation sequence exhibits a kink \citep[e.g.][]{Dungee2022,Bouma2023}. The older the cluster the lower the mass where the kink locates. These three observational features are thought to be related. 

The most accepted explanation for these observations is a mass-dependent core-envelope coupling \citep{Spada2020}. The angular momentum transport between the fast core and the slow envelope during the coupling would result in a temporary stall in the surface spin-down. The observations of the intermediate-rotation gap are consistent with this hypothesis, particularly because it is not observed in fully convective stars \citep{Lu2022}, where a core-envelope coupling would not occur. Nevertheless, to better map this transition and understand the physical processes behind it, more observations are needed namely for stars or clusters with reliable stellar ages.

\subsection{Sun in transition}\label{sec:transition}

The {\it Kepler} observations allowed us to measure the surface rotation of a large sample of {\color{black} MS} low- and intermediate-mass stars \citep[e.g.][]{2014ApJS..211...24M,2021ApJS..255...17S}. These observations also provided the most precise ages for a subsample of stars thanks to asteroseismic studies \citep{2017ApJ...835..173S}. By comparing these ages with those computed from gyrochronology relations -- empirical relations between stellar rotation and age based on young stars and clusters as well as the Sun -- it appeared that the stars of the {\it Kepler} field and older than the Sun were rotating faster than expected \citep[e.g.][]{2015MNRAS.450.1787A,2016Natur.529..181V}.  One explanation for that observation is that the magnetic braking weakens at some point and thus the star is not slowing down as much as those empirical gyrochronology relations were predicting. One theory put forward is that there is a change in the surface differential rotation, the topology of the surface magnetic field, and the magnetic activity cycle \citep{2016ApJ...826L...2M}. However, the origin of such weakened magnetic braking is under debate as the current dynamo models do not reproduce the observations. There is also some discussion about an observational bias against old weakly active stars \citep{2021ApJ...908L..21R}. However, the fact that rotation obtained from asteroseismology, with the opposite bias, also encounters such behavior \citep{2021NatAs...5..707H} seems to confirm the observed trend.
This discovery opens the window to more questions such as: Is the Sun in a transition phase in terms of its magnetic activity? What are the detailed mechanisms in the solar/stellar dynamo? To answer those questions, more observations of the magnetic activity of a large sample of stars, including stars older than the Sun, with precisely known fundamental parameters, as well as long-term observations (many years) of their magnetic activity are required. 
Given that the Sun might be at the edge of going through this transition, it conveys the importance of the solar-stellar connection to better understand the future of the Sun.


\subsection{Angular momentum transport} \label{sec:AMT}

The photometric observations of the CoRoT and {\it Kepler} missions provided constraints on the internal rotation of evolved solar-like stars (subgiants and {\color{black}RG}) thanks to asteroseismic studies. However, stellar evolution models that include treatment of angular momentum transport do not reproduce the observations: those evolved stars{\color{black}'} core rotation is an order of magnitude lower than what is predicted by the models \citep{2012A&A...544L...4E,2013A&A...555A..54C}. This suggests that current stellar models lack some physics. Several candidates are proposed (among which internal gravity waves, magnetic field, the existence of a coupling or not between the core and the envelope of stars), and more high-quality observations of solar-like stars at different evolutionary stages are needed. In parallel, some works on the stellar models are investigating this problem \citep[e.g.][]{2019A&A...621A..66E,2019A&A...631L...6E,2022A&A...664L..16E}. Some of the burning questions regarding this topic are the following: how can we reproduce the observed internal rotation profiles? Is there a fossil magnetic field in solar-like stars? What is the interplay between angular momentum transport and magnetic field? Understanding and implementing the detailed processes of angular momentum transport is crucial as this impacts the chemical mixing \citep{1989ApJ...338..424P} and the estimation of stellar ages that are then used in galacto-archeology or exoplanet fields \citep{2015MNRAS.452.2127S,2018ApJS..239...32P}.


Regarding the search for the signature of an internal magnetic field in {\color{black}MS} solar-like stars to {\color{black}RG}, very exciting discoveries have been made recently. An internal magnetic field can indeed affect the modes, which was studied theoretically \citep[e.g.][]{GooTho1992,2021A&A...650A..53B,2021MNRAS.504.3711L,2021A&A...647A.122M}. The analysis of the aforementioned mixed modes in {\it Kepler} {\color{black}RG} showed the signature of that internal magnetic field \citep{2022Natur.610...43L,2023A&A...670L..16D} providing new constraints for stellar models.

\subsection{Red giant mass loss}

It is known that during their evolution {\color{black}RG} undergo mass loss. This has been measured by comparing masses of red giants at different evolutionary stages (red-giant branch{\color{black}, RGB,} to red clump) \citep[e.g.][]{2012MNRAS.419.2077M, 2016A&A...590A..64S}. However, measurements as a function of different stellar parameters (such as metallicity or luminosity) have not been done, meaning that the detailed mechanisms taking place in the mass-loss process are not well known. Having precise masses determined with asteroseismology is crucial to better understand the mass-loss and how to better implement it in stellar evolution models. The mass-loss in red giants is also important to know the initial masses of stars, which will impact the age determination. Finally, knowing the mass loss on the red-giant branch can affect the initial conditions that are used for AGB (asymptotic giant branch) and post-AGB stars models, which are populations used for extra-galactic studies.

\subsection{Evolution of binary stars}

Binary stars represent a large fraction of the systems in our Galaxy. Their evolution can have different paths depending on the nature of their interactions. Indeed, these binary stars can have tidal interactions or common envelope evolution that can lead to mergers (also called blue stragglers)\citep[e.g.][]{2023AJ....166..154B}. {\color{black}A particular class of interacting binaries, known as Heartbeat stars, comprises eccentric binaries, whose tides lead to distortion and excite pulsations \citep[e.g.][]{Thompson2012,Kolaczek-Szymanski2021}. Tidal interactions can also cause enhanced activity in the components of close-in binaries \citep[e.g.][]{Basri1985}. The investigation of binary systems is thus} important to understand the evolution of cluster stars. The study of some clusters through asteroseismology that provides precise mass measurements gave evidence of stars with masses different than expected \citep[e.g.][]{2011ApJ...729L..10B} that probably resulted from the mass-transfer between stars in a binary system. Asteroseismic analyses of a large number of binary stars in clusters of different ages and metallicities will bring insights into the modelling of interacting binaries. {\color{black} Combined with radial velocity observations and/or {\it Gaia}, this will also contribute to improving our understanding of the evolution of stars in multiple-systems in comparison to single-star evolution. This includes the systems' evolution and associated timescales (e.g. circularization).} This will help us to better estimate event rates of binary interaction, initial-mass-ratio/period distributions, and binary population. We still need to measure the efficiency of the transfer of orbital energy of the inspiraling cores to the envelope during the common-envelope phase of evolution. This is key for many binary systems involving black holes, neutron stars, white dwarfs, type Ia-supernovae, X-ray binaries etc. 


Combining precise asteroseismic measurements with {\it Gaia} recent observations \citep{2021A&A...649A...1G}, can also provide interesting constraints on binary systems formation and evolution. Known binary systems with both astrometric data and asteroseismic properties are increasingly providing more statistics to extract information on binary evolution \citep{2022A&A...667A..31B,2024A&A...682A...7B}.


\section{Priority for the space programs}

Previous missions such as the ESA CoRoT, NASA {\it Kepler} then K2, and now TESS provided a large number of observations allowing a big revolution in stellar physics thanks to asteroseismic studies. However, all those missions were mostly designed for planet search and the sample selection function was biased. For instance, the observations of the {\it Kepler} telescope had a bias towards low-metallicity \citep{Cat2015,Mathur2017}, and not many clusters were observed because of the crowding and difficulties in doing the follow-up for exoplanet confirmation. To better answer the key knowledge gaps listed above, we would need to study simpler populations, with a broad range of stellar parameters in order to understand the impact of each of them on the evolution of stars. 

{\color{black}Having a full comprehension of how the magnetic activity of stars changes over time will greatly influence exoplanet research. This understanding is crucial for enhancing the detection of exoplanets (in particular Earth-like planets around Sun-like stars) through methods like radial velocity measurements or transit photometry and spectroscopy. The presence of spots on stars and their magnetic behavior can sometimes mimic the signals we expect from exoplanets, making it essential to differentiate between the two accurately. Given the current and future programs to search and characterize exoplanets with extreme precision radial velocity instruments as well as space-based transit photometry, studying the magnetic activity of a large sample of stars is of paramount importance for both stellar and exoplanet fields. This will also contribute to our knowledge of how planetary systems and their potential habitability evolve.}



{\color{black} \emph{Photometric observations for asteroseismology}}

{\color{black}Obtaining precise stellar parameters and information on the surface and internal structure and dynamics of solar-like stars from the MS up to the RGB will allow us to tackle all the key knowledge gaps mentioned above.} To take out the most of the seismic information from photometric observations we require a high photometric precision in the brightness measurement. More precise radius, mass, and age can be obtained when individual modes are detected. This requires long and continuous observations of several months, which cannot be done from the ground. {\color{black} The 4 years of the {\it Kepler} observations} showed the importance of asteroseismology in providing precise stellar parameters. {\color{black}\textit{Kepler}, with a projected pixel size of 4'' and a point spread function (PSF) of 21'', collected the best data set for seismic characterization so far. Furthermore, for the detection of modes in MS solar-like stars, a high cadence (at least  1 minute as the modes are above a frequency of 1\,mHz) is required for at least 1 year. This would yield a proper characterization of the modes in MS stars and the detection of the effect of rotation on them. For more evolved subgiants and RG, a cadence of 10 and 30 minutes, respectively, for at least 3 months would be enough to detect and characterize the modes. While continuous observations ease the process, gaped data are still useful \citep{2023A&A...674A.106G}. However, {\it Kepler} observed only one fixed field. TESS overcame this by performing an almost full-sky survey. Given the 27-day sector length and the lower precision of TESS, detection of solar-like oscillations in MS stars has been challenging \citep{2022AJ....163...79H, 2023A&A...669A..67H}. Moreover, the TESS projected pixel size is 20'' (PSF of 84''), making its observations prone to contamination by nearby targets. With the incoming PLATO mission, we expect a step forward with several thousands of solar-like stars with high signal-to-noise seismic detection \citep{2024A&A...683A..78G}, where the noise level is around 50\,ppm h$^{1/2}$, slightly higher than for {\it Kepler}. This would be our reference upper limit for the proposed missions. Nevertheless, one restriction of PLATO is the pixel size of 15'' (PSF of 37''), which prevents resolving stars, particularly in clusters (globular or open).} A white paper proposing the mission HAYDN {\color{black}(High-precision AsteroseismologY of DeNse stellar fields)} was submitted for the ESA Voyage 2050 long-term plan \citep{2021ExA....51..963M}{\color{black}. HAYDN would not only be a dedicated mission to investigate stellar clusters but also the Milky Way's bulge and neighboring dwarf galaxies. Comparatively, with previous missions, HAYDN's projected pixel size would be 1'' with a PSF of 1.3'', which was estimated to be the minimum requirement for avoiding contamination by nearby sources in crowded fields.} While selected for phase 0 of the ESA M7 call, it was not selected for phase A. HAYDN will be proposed {\color{black}in} the next ESA M8 call. {\color{black}Nevertheless,} having observations in the meantime would prevent long periods of time without data. 

{\color{black} \emph{EUV and X-rays observations}}

The study of stellar magnetic activity requires long-term observations{\color{black}. In particular, they would allow the detection of} full cycles that could be longer than a decade for a star like the Sun (and even longer for more evolved and hotter stars), {\color{black} improve our knowledge of the processes governing solar and stellar dynamo.} A few hundred stars have been monitored from the ground but having both magnetic activit{\color{black}y and} precise stellar parameters from asteroseismology will provide key information to understand the detailed mechanisms involved in the magnetic activity of the Sun and stars. Studying magnetic activity can be done in different wavelengths. One very useful wavelength is the {\color{black}EUV} as it allows us to study flares and coronal mass ejections in other stars. So far, the Extreme Ultraviolet Explorer (EUVE) was the only observatory that has extensively done spectroscopic observations in that wavelength. No other observations will be done in the future and this will bring a gap in our study of stellar magnetism. So, it is very critical to have such a telescope planned in the future. Observations in X-rays can also provide another way of studying the magnetic activity of many stars. 
{\color{black} Such observations would help us address the key knowledge gaps on the intermediate-rotation gap (Sect. \ref{sec:rotgap}), the Sun in transition (Sect. \ref{sec:transition}) and angular momentum transport (Sect. \ref{sec:AMT}).}

{\color{black} \emph{Stellar imager}}

Another phenomenon related to stellar magnetic activity is the presence of spots that are not well studied in stars other than the Sun. From their evolution, lifetime, differential rotation, active latitudes, very little is known on starspots, except for the information from spectropolarimetry. Being able to resolve stellar surfaces can allow us to answer many questions on stellar magnetism. {\color{black} For instance, obtaining information on active latitudes and differential rotation can provide hints to investigate the key knowledge gaps on the intermediate-rotation gap (Sect. \ref{sec:rotgap}), the Sun in transition (Sect. \ref{sec:transition}) and the angular momentum transport (Sect. \ref{sec:AMT}) by better constraining dynamo models. Observations over at least a decade would be useful to study the evolution of spots over a magnetic cycle, such as the 11-yr cycle of the Sun.} An idea, {\color{black} the Stellar Imager \citep{2010arXiv1011.5214C},} was suggested {\color{black} and proposed to NASA} that involved a UV-optical Fizeau interferometer with 20 or 30 one-meter mirrors in a flying formation. The length of the interferometer could be up to 10 km. Being able to build such an interferometer on the Moon (and maybe Mars) would be an amazing opportunity.

{\color{black} \emph{Multi-wavelength simultaneous observations for stellar magnetic activity}}

The magnetically-driven variability of Sun-like stars is caused by dark spots and bright facular regions transiting stellar disks as stars rotate. White-light observations (e.g. performed by {\it Kepler} and TESS and planned for PLATO) do not allow distinguishing between variations brought by facular regions and by spots \citep{2016A&A...589A..46S,2024ApJ...963..102L}. This significantly {\color{black}hinders} 
the determination of rotation periods of slow rotators like the Sun since the interplay between spot and facular contributions to brightness variations of such stars causes irregularities in their light curves. The difficulties in detecting periods of slowly rotating stars might be an important contributor to the explanation of the lower-than-expected number of observed G-type stars with near-solar rotation periods and also hampers the solar-stellar comparison studies \citep{2021ApJ...908L..21R,2021ApJS..255...17S}. Monitoring of stellar brightness in several spectral passbands, i.e. multi-color photometry will circumvent this limitation of white-light observations. Indeed, the monochromatic facular contrast decreases strongly with the wavelength, while the spot contrast does not show such a pronounced dependence \citep{2024ApJ...963..102L}.  As a result, stellar brightness variability is expected to be faculae-dominated in the UV and spot-dominated in the visible spectral domain (e.g. solar rotational variability is faculae-dominated shortward of 400 nm and spot-dominated longward of 400 nm). Consequently, simultaneous monitoring of stellar brightness in the UV and visible spectral domains will allow separating spot and facular contributions to stellar brightness variations. Not only will this lead to a more reliable determination of stellar rotation periods but also to a better understanding of stellar magnetic activity in general, {\color{black} providing answers for the key knowledge gaps on the core-envelope coupling (Sect.  \ref{sec:rotgap}), the Sun in transition (Sect. \ref{sec:transition}), and angular moment transport evolution (Sect. \ref{sec:AMT})}. 
Furthermore, stellar intrinsic variability depends on wavelengths in a different way than photometric signatures of the planetary transits. Consequently, multi-color photometry can facilitate the {\color{black} distinction} between intrinsic variability and planetary transits, which would lead to a more reliable detection and characterization of exoplanets.

{\color{black} The proposed missions to study stellar magnetic activity would require several years of observations spanning at least a decade to study cycles and a cadence of at least 4 hours.}

All the proposed experiments are summarized in Table~\ref{tab1}.

\begin{table}
    \centering
    \begin{tabular}{l|c|c|l}
         & & Related recent& \\
         Open fundamental & Focus of the & and future space & \\
         scientific question & ESA experiment& experiments & Short, Middle or long term\\\hline\hline
         Is the the Sun in a & Moon, & CoRoT, & Middle: {\color{black} photometer (UV+vis.) } \\
         transition phase in & Mars, & {\color{black} {\it Kepler}}, & {\color{black} $>$\,1yr obs., $\sim$\,1min cad.} \\
         terms of its magnetic & BLEO & TESS, & {\color{black} EUV and X-rays}\\
         activity? &  & PLATO  & {\color{black} Simultaneous to photometric}\\
         &  &  & Long: an instrument similar\\
         &  &  & to the proposed Stellar imager\\
         &  &  & that requires more technical\\
         &  &  & and engineering developments.\\
         &  &  & {\color{black}$>$\,10yr obs.}\\\hline
         How does mass loss & Moon, & {\color{black} {\it Kepler}}, K2, & Middle: photometer {\color{black} (UV+vis.)}\\
         operate in red giants?& Mars, & TESS, & {\color{black}$>$\,1yr obs.}\\
         & BLEO & PLATO & {\color{black} $\sim$\,30 min cad.}\\
         & & {\color{black} Roman}\\\hline
         How does the angular & Moon, & CoRoT, & Middle: photometer {\color{black} (UV+vis.)}\\
         momentum transport & Mars, & {\color{black} {\it Kepler}}, & {\color{black} EUV and X-rays}\\
           evolve in low-mass stars& BLEO & PLATO & {\color{black}$>$\,2yr obs.} \\
            {\color{black} from MS to RG?}& & & {\color{black} $\sim$\,1min cad.}.\\\hline
         How do binary systems & Moon, Mars, & {\color{black} {\it Kepler}}, K2, & Middle: photometer {\color{black} (visible)}.\\
         evolve? & BLEO
         & TESS, PLATO & {\color{black}$>$\,1 yr obs.}\\
         & & &{\color{black} $\sim$\,30 min cad.}\\\hline\hline
    \end{tabular}
    \caption{Recommendations for addressing key questions in stellar physics with the ESA HRE programme in the short, middle, and long term.}
    \label{tab1}
\end{table}


\section{Future outlook and summary}
We described the different recommendations of future possible experiments within the ESA's Human Robotic Exploration programme in order to fill key knowledge gaps and answer key questions in stellar physics. These experiments can be set up in low Earth orbit and on the Moon or Mars and mostly within 10 years from now. Combined with already planned missions, the HRE programme will tremendously contribute to a leap forward in our understanding of how stars' dynamics and structure evolve, impacting other fields such as exoplanet characterization and galacto-archaeology.

\bmhead{Acknowledgments}

{\color{black} The authors thank the European Space Agency for the opportunity to contribute this perspective article, based on the Stellar Physics section of the Astrophysics SciSpacE White paper \citep{ESASciSpacewhite21}.}
S.M.\ acknowledges support by the Spanish Ministry of Science and Innovation with the Ramon y Cajal fellowship number RYC-2015-17697, the grant number PID2019-107187GB-I00, and the grant no. PID2019-107061GB-C66.
A.R.G.S acknowledges the support from the FCT through national funds and FEDER through COMPETE2020 (UIDB/04434/2020,  UIDP/04434/2020 \& 2022.03993.PTDC) and the support from the FCT through the work contract No. 2020.02480.CEECIND/CP1631/CT0001.

{\color{black}
\bmhead{Data availability}

Data sharing is not applicable to this article as no datasets were generated or analysed during the current study.

\bmhead{Competing Interests}
The authors declare no competing interests.

\bmhead{Author Contributions}
S. M. contributed to the ESA science community white paper related to this topic. 
S. M. and A. R. G. S. adapted and extended the text to form this perspective paper.
}









\bibliography{BIBLIO_sav,moreRefs}

\end{document}